\documentclass[%
 reprint,
 amsmath,amssymb,
 aps,
]{revtex4-2}

\usepackage{braket}
\usepackage{breqn}
\usepackage{gensymb}
\usepackage{graphicx}
\usepackage{dcolumn}
\usepackage{bm}
\usepackage{subfigure}

\begin{document}

\preprint{APS/123-QED}

\title{Vector-based feedback of continuous wave radiofrequency compression cavity for ultrafast electron diffraction}
\author{Thomas M. Sutter}
\email{Corresponding Author: tsutter@ucla.edu}
\author{Joshua S. H.  Lee}
\author{Atharva V. Kulkarni}
\author{Pietro Musumeci}
\author{Anshul Kogar}

\affiliation{Department of Physics and Astronomy$,$ University of California Los Angeles$,$ Los Angeles$,$ CA 90095$,$ USA}


\date{\today}

\begin{abstract}
The temporal resolution of ultrafast electron diffraction (UED) at weakly relativistic beam energies ($\lesssim$ 100 keV) suffers from space-charge induced electron pulse broadening. We describe the implementation of a radio frequency (RF) cavity operating in the continuous wave regime to compress high repetition rate electron bunches from a 40.4 kV DC photoinjector for ultrafast electron diffraction applications. Active stabilization of the RF amplitude and phase through a feedback loop based on the demodulated in-phase and quadrature components of the RF signal is demonstrated. This scheme yields 144$\pm$19~fs RMS temporal resolution in pump-probe studies. 
\end{abstract}

\maketitle


\section{Introduction}

Ultrafast electron diffraction (UED) probes the time-dependent spatial charge distribution of a many-particle system in response to an impulsive excitation~\cite{mourou1982, zewail2006, millersciaini}. Typically, a laser pulse is used to excite a sample away from equilibrium and a subsequent pulse of electrons diffracts from the perturbed system. In solids, this experimental scheme provides snapshots of the non-equilibrium crystal structure. Varying the relative arrival time between the pump and probe pulses allows for a full mapping of the structural dynamics in response to photo-excitation. In a UED experiment, a crucial parameter is the temporal resolution; this is affected by the temporal widths of the pump laser and electron pulses in addition to the jitter in their relative arrival time. 

Compact ultrafast electron diffraction instruments have greatly improved in their temporal resolution and beam quality in the last decade due to the adoption of methods from the field of accelerator and beam physics~\cite{filippetto:rmp}. Some of the important recent developments include the widespread use of radio frequency (RF)~\cite{vanoudheusden:JAP,vanoudheusden:PRL} and magnetic compression schemes~\cite{qi2020breaking, kim2020towards}, the use of high brightness photo-cathodes~\cite{maxson:ued}, and the development of time-stamping diagnostics~\cite{scoby2010, zhao2018terahertz}. Each have contributed to improving the UED instruments by enhancing spatial resolution as well as temporal resolution to below the 100~fs regime. While beamlines at highly relativistic energies based on RF injectors still require significant infrastructure investment and are mainly restricted to national laboratory settings~\cite{weathersby:SLACUED, siddiqui:LBNL, zhu:BNL}, lower energy 25-100 keV scale beamlines based on DC photoinjectors are well-suited to a university-sized laboratory~\cite{siwick:RFcompression, centurion:ued, maxson:ued}. Lower beam energy instruments can take full advantage of the development of high repetition rate ultrafast laser systems. These lasers are now available with sufficient energy per pulse to perform pump-probe experiments at repetition rates of 10 kHz and above, up to the limit of sample recovery times. 

At weakly relativistic beam energies, space-charge effects (i.e. Coulomb repulsion) significantly broadens the electron bunch longitudinal width resulting in a severe degradation of the temporal resolution in UED experiments. Electron bunch compression (e.g. using RF cavities) is essential to obtaining short pulse lengths. In order to achieve maximum compression, the electron pulse is typically injected into the RF cavity at the zero-crossing of the field (i.e. at the point where the electron pulse experiences zero net change in momentum). This strategy ensures that a nearly linear energy chirp is imparted onto the beam longitudinal phase space, which leads to a strong compression of the electron pulse as it drifts towards the sample~\cite{vanoudheusden:JAP, vanoudheusden:PRL}. 

Unfortunately, such a configuration also implies that any small timing error at the entrance of the cavity translates into a variation in the electron beam energy, ultimately degrading temporal resolution. Synchronization between the laser pulse that generates the electron bunch and the field in the RF cavity is therefore crucially important. 
The phase of the RF field is a key parameter in this setup, yet it is difficult to fully control. For example, due to the narrow frequency resonance of typical RF compression cavities, small changes in temperature cause significant phase shifts. Similarly, stability in the amplitude of the RF field is also important for optimal beam compression and arrival time stability. Earlier efforts in stabilizing the time-of-arrival jitter have concentrated on phase feedback~\cite{siwick:jitter}. However, a combined phase and amplitude vectorial feedback could further improve the control of the RF parameters and lead to new applications in electron pulse manipulation with RF cavities.

In RF circuits, IQ (in-phase/quadrature) double balanced mixers enable simultaneous acquisition of the amplitude and phase of an RF signal with respect to a local oscillator reference. Notably, the mixers also work in reverse; by providing two separate voltage inputs at the I and Q ports, one can modulate the phase and amplitude of an input RF signal. Thus, adding IQ mixers upstream and downstream of the RF cavity allows for control and measurement of the RF field's phase and amplitude. Using these mixers in combination, we can perform feedback control with a bandwidth mainly limited by the processing time of the PID feedback loop which minimizes the effects of long term drifts in the system.

In this paper, we discuss the design and implementation of the GARUDA (Garuda Ultrafast Diffraction Apparatus) beamline: a novel low energy (40.4 keV) UED beamline at UCLA based on the use of an RF cavity operating in continuous wave (CW) mode. We utilize a system of double-balanced IQ mixers to provide simultaneous phase and amplitude feedback that enables reaching an instrument response function of less than 150~fs RMS, as measured in a pump-probe study of ultrafast melting of a charge density wave superlattice. It is important to note that such a temporal resolution measurement is achieved over several hours, during which the drift in time-zero is minimized by the implementation of the feedback loop. The main residual contribution to the temporal resolution of the instrument is due to the electron bunch length.

\section{RF-based bunch compression for UED beamline}

Near the zero-crossing of the RF field, the momentum change imparted onto electrons propagating through a cavity depends almost linearly on the electron's position within the bunch.
Such RF cavities can therefore be described as temporal lenses~\cite{pasmans:microwave}.
Propagation in a drift space will transform the momentum-position correlation and flatten the longitudinal phase space at the temporal focus. 

The effective focal length for a cavity 
is the distance over which an energy collimated beam injected into the cavity will reach its minimum longitudinal length ~\cite{denham2021analytical} and can be written as:
\begin{equation}
f = \frac{m_0c^2\gamma^3\beta^3}{e_0V_ck\cos\psi}
\end{equation}
where $m_0$ and $e_0$ are the rest mass and charge of the electron respectively, $k = 2\pi/\lambda$ is the RF wavenumber, $\gamma$ and $\beta$ the standard relativistic factors, respectively, and $\psi$ is the RF phase deviation from the zero-crossing condition (which is defined as the phase where the average momentum of the beam does not change when passing through the cavity). For electrons of kinetic energy $E_k$ = 40.4~keV ($\beta$ = 0.376), RF frequency $c/\lambda$ = 2.856~GHz and a cavity with accelerating voltage $V_c = 3.04$~kV, the focal length of the temporal compression lens can be calculated to be 0.187~m. Note that a longitudinally diverging beam will come to a temporal focus some distance after this cavity focal length.

For small fluctuations of the RF amplitude and phase, the focal length changes and the electron pulses are not optimally compressed at the sample plane. These fluctuations also alter the beam energy which changes the time-of-arrival of the electrons. This latter effect is a significantly more important contribution to the deterioration of the temporal resolution in a UED measurement. The change in electron arrival time compared to the reference case is given by,
\begin{equation}
t_a = \frac{D/c}{\beta^3\gamma^2} \frac{\Delta \gamma}{\gamma}
\end{equation}
where $D$ is the propagation distance from the cavity to the sample and 
\begin{equation}
\frac{\Delta \gamma}{\gamma} = \frac{e_0 V_c \sin \psi}{m_0c^2 \gamma} 
\end{equation}
depends on the cavity voltage and the injection phase. If the cavity is tuned exactly to the zero-crossing phase ($\psi = 0$), fluctuations in RF amplitude have no effect on the electron arrival time. However, if there is a small error in phase, the time of arrival will also change due to fluctuations in the RF amplitude. Thus, combined phase and amplitude jitter introduces an additional 2nd order jitter in the electron arrival time. In our numerical example, for a distance $D = 0.279$~m at the zero-crossing condition, we expect
\begin{equation}
\frac{dt_a}{d\psi}\Big\rvert_{\psi=0}  = \frac{D e_0 V_c}{m_0 c^3 \beta^3 \gamma^3}
\label{dtdpsi}
\end{equation}
which yields $\frac{dt_a}{d\psi} = $ 1.47 ps/degree. If we are 1$^\circ$ off from the zero-crossing condition, a 2.5\% amplitude change will lead to 40 fs time-of-arrival difference.

\subsection{Beamline design}

\begin{figure*}
    \centering
    \includegraphics[scale=0.4]{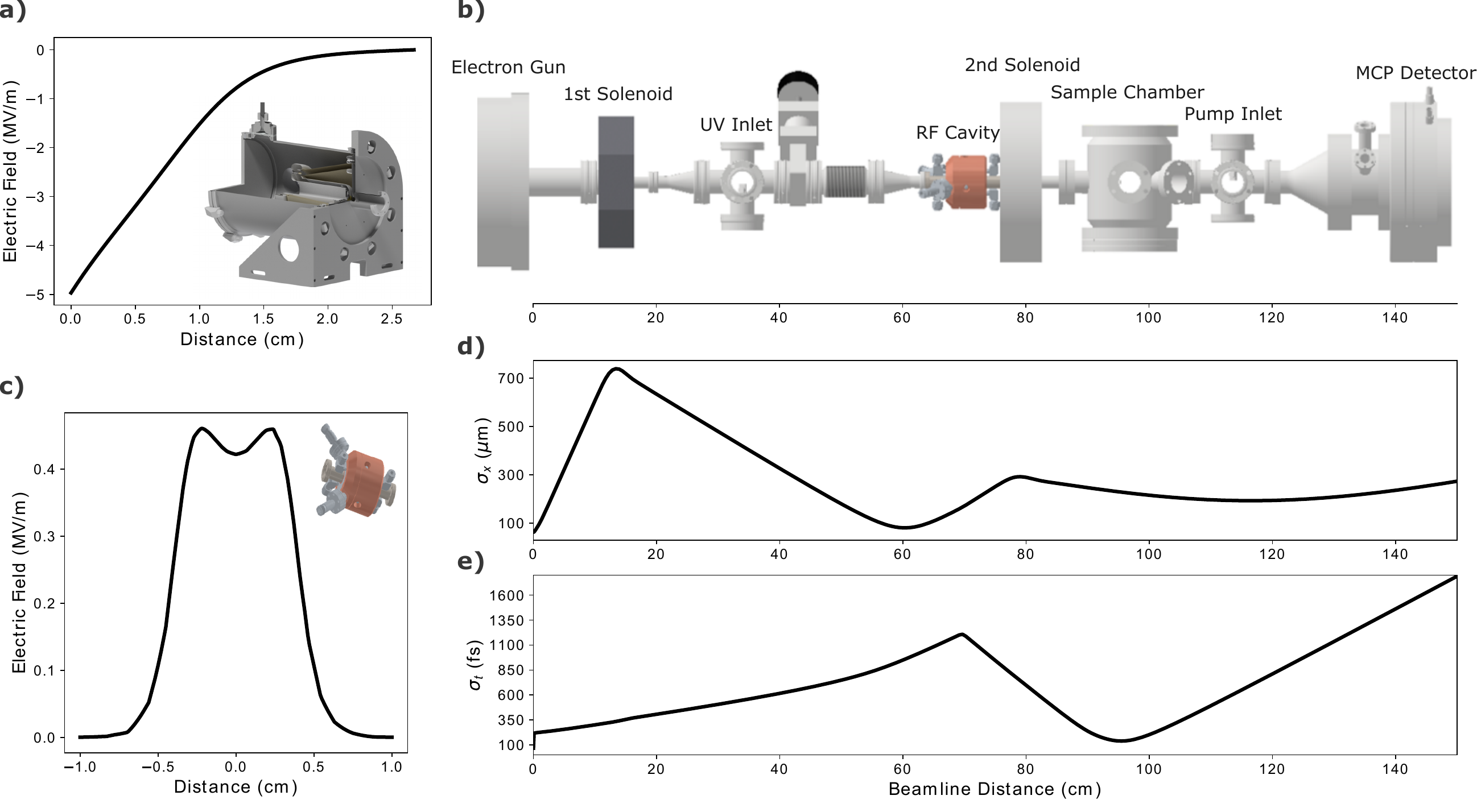}
    \caption{(a) Map of the electric field of the 40.4 kV DC electron gun (From DrX Works\textsuperscript{TM}) along the central axis. (b) CAD model of the GARUDA beamline. Beam steering optics are omitted from this drawing. (c) Map of the peak electric field along the central axis of the RF cavity. The final figures (d) and (e) show the simulated electron bunch transverse and longitudinal RMS size respectively against the average bunch position. Note that the 100 $\mu$m pinhole is not present in this simulation.} 
    \label{fig:model}
\end{figure*}

Fig.~\ref{fig:model}(b) shows a schematic of the GARUDA electron beamline. It should be noted that four magnetic beam steerers are present on the beamline, but they are not depicted because their exact locations are not critical. Roughly, these steerers are placed (i) directly after the electron gun, (ii) directly after the first solenoid, (iii) before the RF cavity, and (iv) directly after the second solenoid. In our setup, the pump and probe pulses originate from a 1030~nm, 180~fs FWHM Yb-based Pharos\textsuperscript{TM} laser system  with 10~W of average power and repetition rate up to 20~kHz. A pulse picker allows for tuning the repetition rate from the regenerative amplifier; for this study, we operate at 500 Hz. The output is routed to a beam-splitter that separates the pulses into pump and probe paths. The pump is directed into an optical parametric amplifier (OPA) to obtain a wavelength-tunable output in the optical-to-infrared range of the electromagnetic spectrum. This laser pulse is then free-space coupled into a vacuum chamber housing the sample to provide an impulsive excitation that drives the sample away from equilibrium. On the probe path, 1030~nm light is frequency-doubled twice via nonlinear crystals to provide 257.5~nm (4.8~eV) laser pulses. These pulses are then focused to a 60~$\mu$m FWHM spot on a poly-crystalline copper cathode housed in vacuum. An electron bunch is thus produced via the photoelectric effect. Increasing the bunch charge provides an improved diffraction signal-to-noise at the cost of greater transverse and longitudinal broadening. The compromise between these effects must be weighed carefully in selecting an operating bunch charge. For the GARUDA beamline, the operating bunch charge may be easily varied depending on the requirements of a specific study (here we present data taken at 1 fC and 3.3 fC).


The electron bunch is then accelerated by a DC field to 40.4~keV kinetic energy over the cathode-to-anode distance of $\sim$11~mm. This brings their velocity (normalized to the speed of light) to $\beta = 0.376$. Initially, the transverse and longitudinal size of the electron pulse grows rapidly due to the non-zero thermal emittance and the mutual Coulomb repulsion (space-charge). A solenoid magnetic lens directly after the gun is used to control the transverse profile of the beam and focus it into the entrance of the RF compression cavity. At the appropriate RF cavity phase, the electron pulse is given a negative chirp with no net bunch acceleration. Consequently, the mean velocity of the beam remains constant, but electrons at the back (front) of the pulse are accelerated (decelerated). Thus, the pulse will come to a longitudinal focus at some distance after the RF compression cavity. At the ideal RF power and phase, this focus will be at the sample plane. In our setup, this distance is 27.9~cm downstream of the RF cavity.

When the cavity is set at the longitudinally focusing phase, the RF field has a defocusing effect on the transverse beam profile. A second solenoid refocuses the beam to a transverse size ($\sigma_x$) of approximately 200~$\mu$m RMS at the sample plane. This is illustrated in Fig.~\ref{fig:model}(d). Here there is a trade-off between spot-size on the sample and $\textbf{q}$-space resolution on the detector. In order to achieve a smaller spot-size at the sample plane, a retractable 100~$\mu$m diameter pinhole is placed directly before the sample. Control over the spot size in this way comes at the cost of decreased intensity (typical transmission through the pinhole is 20\%).

\subsection{Beamline Simulation}

We utilize the General Particle Tracer (GPT) software package to simulate the propagation of electron bunches along the beamline. GPT is a three-dimensional particle tracking software that numerically simulates charged particle dynamics in external electromagnetic fields while accounting for space-charge effects~\cite{GPT}. All of the GPT results presented here are from simulations with a 40.4~keV beam energy, 1.0 fC bunch charge, and 0.25~eV thermal emittance. Note that the 100~$\mu$m diameter pinhole is not included in the simulation.

\begin{figure}
    \centering
    \includegraphics[scale=0.5]{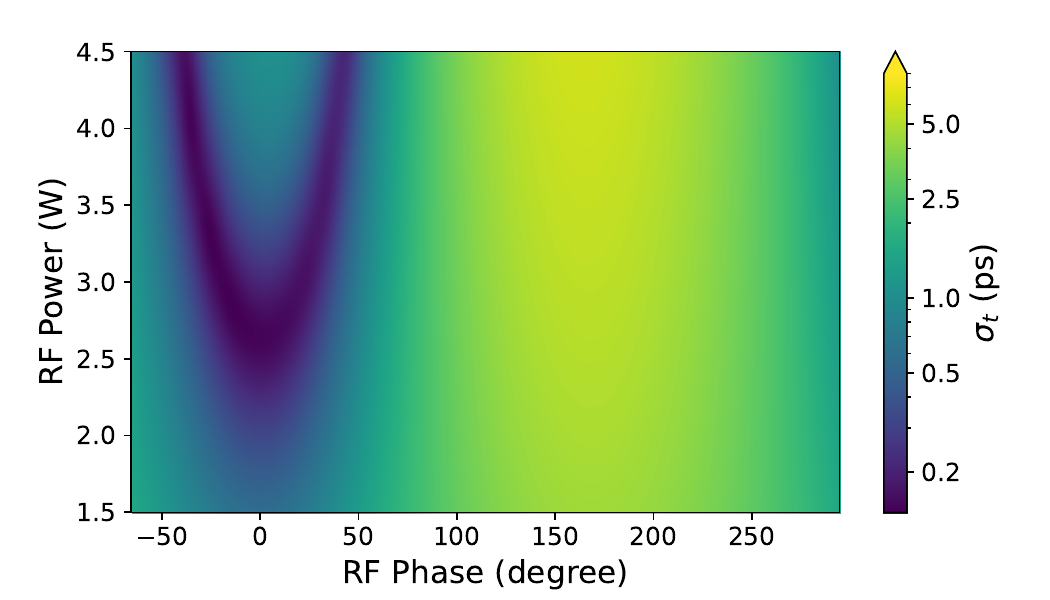}
    \caption{Simulated temporal RMS width ($\sigma_t$) of the electron pulse at the sample plane as a function of RF phase and RF power. The compression reaches $\sigma_t \approx 140$~fs. The simulations were performed with 40.4~keV beam energy, 1.0 fC bunch charge, and 0.25~eV thermal emittance.}
    \label{fig:gpt}
\end{figure}

Fig.~\ref{fig:model}(c) and (d) show the simulated transverse ($\sigma_x$) and temporal ($\sigma_t$) bunch RMS widths as a function of the average position along the beamline. In this simulation, the RF cavity phase and power are optimized to longitudinally focus at the sample plane while imparting no net momentum to the electron bunch; we will refer to these optimized values as the ideal RF phase and RF amplitude. If we were to vary the RF phase away from this ideal value, $\sigma_t$ would grow until it reaches a maximum expansion phase at 90$^\circ$ away from the compression phase. In contrast, if we were to vary the RF power while keeping phase fixed at the zero-crossing point, the electron bunch would come to a longitudinal focus either before or after the sample plane. Fig. \ref{fig:gpt} shows a simulated mapping of $\sigma_t$ at the sample plane as a function of RF amplitude and phase. For fixed RF powers below the ideal level, the minimizing phase always corresponds to the zero-field crossing, where the field slope is maximal. However, for RF power levels higher than ideal, the minimizing phase along cuts of constant RF power bifurcates into two branches. In these branches, either a net acceleration or deceleration is applied to the electron bunch along with the longitudinal compression; the shift away from the zero-field crossing exposes the electron bunch to a weaker field slope which compensates for the excess power thus holding the focus at the sample plane.

These two bifurcated branches achieve similar simulated $\sigma_t$ compared to the ideal condition; however, arrival time jitter due to amplitude fluctuations is significantly worse along the branches. At the ideal condition, the zero field crossing completely eliminates first order arrival time jitter effects due to fluctuations in the RF power; however, in the high power branches, this arrival time jitter increases more than tenfold to a level of 40 fs per 0.1 \% power variation in the cavity. Additionally, the dependence of $\sigma_t$ on RF phase is more severe in the high power branches compared to the ideal condition. Hence, operating in these branches is undesirable unless amplitude is highly stabilized.

\subsection{RF cavity and cooling}
\begin{figure}
\centering
\includegraphics[scale=0.5]{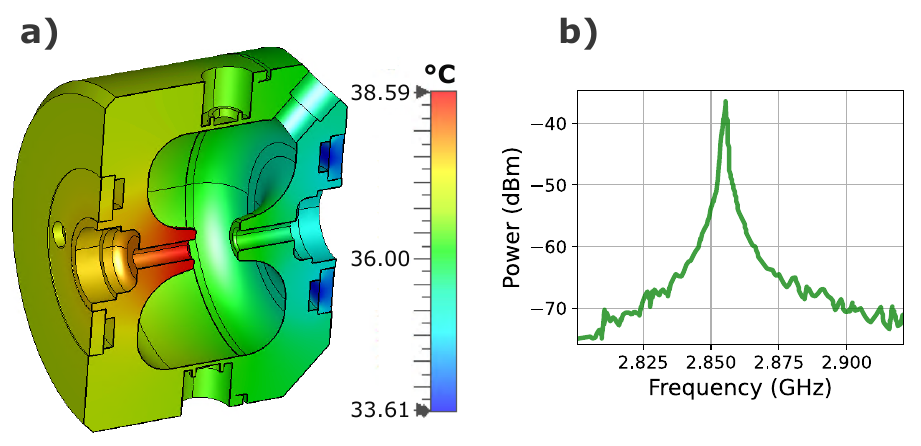}
\caption{(a) Cavity temperature simulation under 50W continuous-wave RF power operating with one cooling channel. The maximum and minimum temperature differ by about 5 $^\circ$C (image courtesy of Radiabeam Technologies). (b) S21 measurements for the cavity conducted in vacuum. The resonance occurs at 2.856 GHz frequency with a bandwidth of 1.4 MHz FWHM.}
\label{fig:cavity}
\end{figure}

A single-cell S-band 2.856~GHz buncher cavity is used to impart the required chirp on the electron bunch. The cavity was constructed by Radiabeam technologies (Fig.~\ref{fig:cavity}) and is based on a re-entrant nose cone geometry. The conversion of RF cavity power ($P$) to the peak accelerating voltage ($V_c$) is quantified by the shunt impedance ($R_s$) through the relation:
\begin{equation}
    R_s = \frac{V_{c}^2}{P}
    \label{eq:shuntimp}
\end{equation}
The re-entrant geometry is designed to obtain a cavity shunt impedance of 3.3 M$\Omega$ so that $<$3~W of RF power is sufficient to reach the nominal cavity voltage of $\sim$3~kV. The distance between the nose cones is carefully chosen so that 40.4~keV electrons will enter and exit the cavity in half a cycle of the RF fields in order to maximize the transit time factor.

The cavity is equipped with two n-type ports, one of which is used to feed the input coupler into the cavity and the other is used for a probe antenna. This antenna is calibrated to a -12~dB coupling in order to monitor the amplitude and phase of the RF fields. 
%

When operating in CW mode, the maximum power into the cavity is limited by heat transfer considerations. Two separate water cooling circuits can be used to control the temperature of the copper structure, but due to a leaky connection, we have operated using only the upstream water channels (Fig.~\ref{fig:cavity}(a)). Although not ideal, heat load simulations indicate this mode of operation is acceptable as it yields a relatively modest peak temperature increase of less than 5$^\circ$C under 50~W of input power. In our implementation, the operating input power is less than 5~W, so this temperature gradient is much smaller. During operation, the temperature of the water in the cooling line is actively stabilized such that fluctuations do not exceed 0.1$^\circ$C.

The cavity dimensions are designed such that the resonant frequency is 2.856 GHz to very high accuracy ($\pm 0.05$ MHz). Mismatches between the cavity resonant frequency and the frequency from the oscillator lead to reflections and losses. The input coupler that converts the RF power from the coaxial cable to the cavity mode has a coupling factor of 0.98 so that the return loss is very small (-40 dB of the input power), as confirmed by the measurement shown in Fig.~\ref{fig:cavity}(b). After tuning and adjustments, a sharp cavity resonance with an unloaded quality factor of 12000 was measured with a vector network analyzer at 2.856 GHz prior to installation on the beamline (Fig.~\ref{fig:cavity}(b)).

\begin{figure}
    \centering
    \includegraphics[scale=0.5]{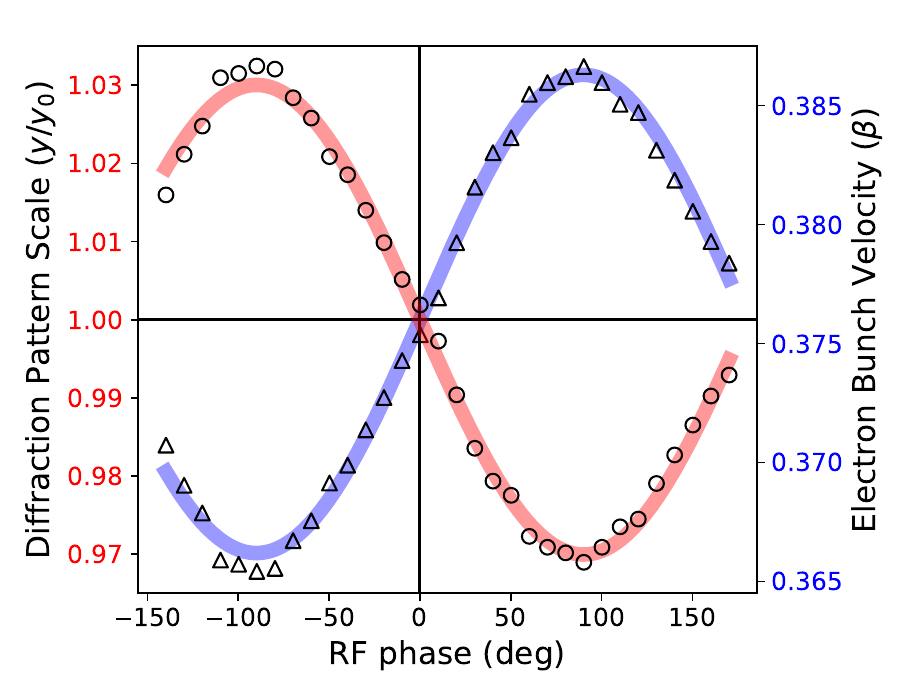}
    \caption{Relative size of an electron diffraction pattern from a thin-film gold crystal as a function of the RF phase in the cavity at 2.3 W input power. This is given by the circular data points and red fit curve, and was directly measured from the distances between several Bragg peaks on the detector. The triangular data points and blue fit curve give the electron bunch velocity inferred from the variation in diffraction pattern scale.}
    \label{fig:deltaQ}
\end{figure}

The design shunt impedance is verified by measuring the change in the beam velocity as a function of the RF phase, as shown in Fig.~\ref{fig:deltaQ}. The deviation in beam momentum is obtained by monitoring the shift in the Bragg peaks from a gold reference sample as a function of the input phase while maintaining an input power of 2.3~W. Acceleration (deceleration) of the electron bunch contracts (expands) the scale of the diffraction pattern on the detector. A sinusoidal fit of this diffraction scale variation with RF phase yields a shunt impedance of 3.2~M$\Omega$, in good agreement with the predictions. Note that this measurement also allows one to easily determine the phase for which no net impulse is imparted to the bunch as described later.


\section{Synchronization and feedback}

\begin{figure*}[ht]
    \centering
    \includegraphics[scale=0.45]{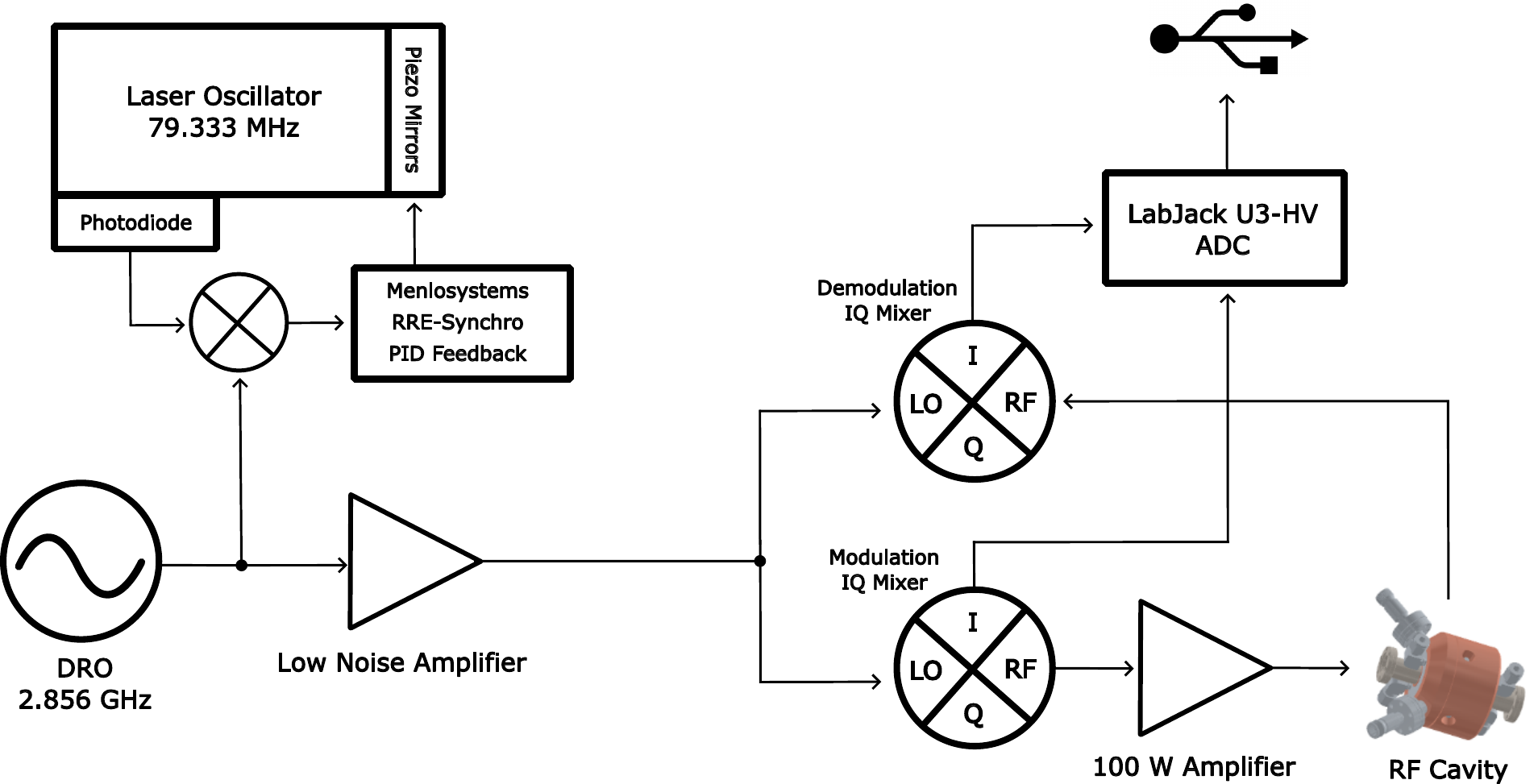}
    \caption{Schematic diagram of the RF compression circuit. This illustrates how the 2.856 GHz RF wave out of the DRO is used to simultaneously synchronize the RF cavity and the laser oscillator. The arrows represent the signal direction.}
    \label{fig:system}
\end{figure*}

RF compression requires precise synchronization between the electron arrival time at the RF cavity and the phase of the RF field inside the cavity. This synchronization problem is central to state-of-the-art RF technologies at large-scale particle accelerators; thus, there is a large body of knowledge in this domain \cite{akre2007lcls,filippetto2022feedback,branlard2012european,zhang2021precision,piersanti2023commissioning}. In the case of UED experiments at the GARUDA beamline, this problem is addressed by synchronizing the laser oscillator (seed for pump and probe pulses) and the RF cavity wave to a master clock.

\begin{figure}[h]
    \centering
    \includegraphics[scale=0.5]{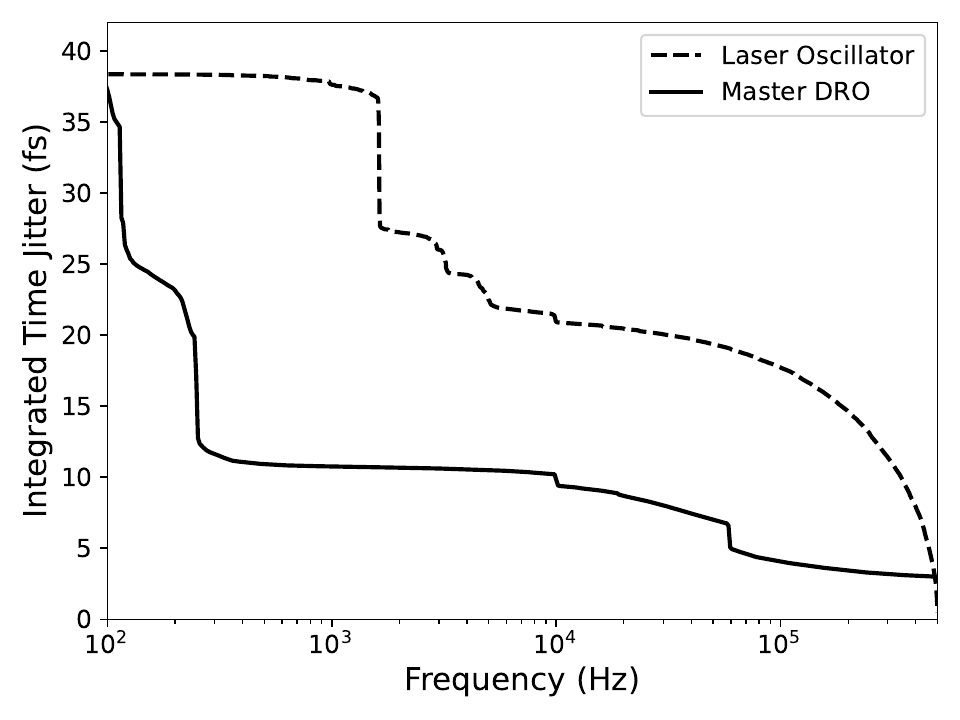}
    \caption{RMS integrated timing jitter of the master dielectric resonator oscillator (2.856 GHz) and the laser oscillator (79.333 MHz).}
    \label{fig:ampphasejitter}
\end{figure}

Fig.~\ref{fig:system} shows a simple schematic of this system. In order to stabilize the RF compression process, the laser oscillator is locked via PID feedback to a master dielectric resonator oscillator (DRO) providing a low noise electronic signal at 2.856 GHz frequency. The frequency of the laser oscillator is 79.333 MHz; pulses from this oscillator are fed into a mixer detector unit [MenloSystems, MDU-2856MHz-FS]. This converts the laser pulses into an electronic signal via a fast photodiode with a bandwidth large enough to resolve the harmonics of the 79.333 MHz pulse train up to the 36th order (2.856 GHz). The DRO signal is mixed with this photodiode signal to produce an error voltage that measures the phase mismatch. Two piezo mirrors are controlled inside the laser oscillator to adjust the length of the cavity and synchronize the phase. The integrated timing jitters of the master DRO and the laser oscillator are shown in Fig.~\ref{fig:ampphasejitter}; both stay below 40 fs RMS.

The RF cavity is driven by an amplified and phase-shifted signal from this same master DRO. The RF amplifier is a fixed gain Empower model 2193 which is run below saturation to allow for amplitude feedback adjustments. The variable attenuation and phase shift is PID feedback controlled by two IQ mixers [Marki Microwave, EVAL-MMIQ-0205H]. These are four port devices (two RF ports and two low frequency ports) which enable simultaneous in-phase and quadrature mixing.

If the following two signals are applied to the RF ports,
\begin{align}
    V_1 &= a \cos{(\omega t)} \\
    V_2 &= b \cos{(\omega t + \phi)}
\end{align}
then the IQ mixer will produce two DC voltages, I and Q, such that:
\begin{align}
    I &= (a/b) \cos{(\phi)} \\
    Q &= (a/b) \sin{(\phi)}
\end{align}
Here, we take $V_1$ as a signal from the master DRO and $V_2$ as a signal from a antenna probe in the RF cavity. The mixer acts to demodulate the signal to the $I$ and $Q$ voltages which yield the relative amplitude and phase of the RF cavity wave. Letting $A = a/b$, the relevant relations are:
\begin{align}
    A &= \sqrt{I^2 + Q^2} \label{eq:amp} \\
    \phi &= \arctan{\left(\frac{Q}{I}\right)} \label{eq:phi} 
\end{align}
Likewise, we can use a separate IQ mixer to set the amplitude and phase for an RF wave applied to port $V_2$. In this case, the $V_1$ signal is modulated according to the DC voltages externally applied at the $I$ and $Q$ ports of the mixer.  With simultaneous operation of two IQ mixers, one acting as a modulator and one acting as a demodulator, we can measure and control the phase and amplitude of the RF field inside the cavity.

Prior to deployment in the feedback system, the IQ mixers must be calibrated. This is accomplished using a 2-port vector network analyzer (VNA) and an analog phase shifter. Ideally, the relationship between the true phase (as measured by the VNA) and the IQ mixers set-phases would be a line with unity slope. In practice, there can be a deviation from this linearity due to a variety of possible causes including the response of the antenna inside the cavity and the actual response of the mixers themselves. 



\begin{figure*}[ht]
    \centering
    \includegraphics[scale=0.8]{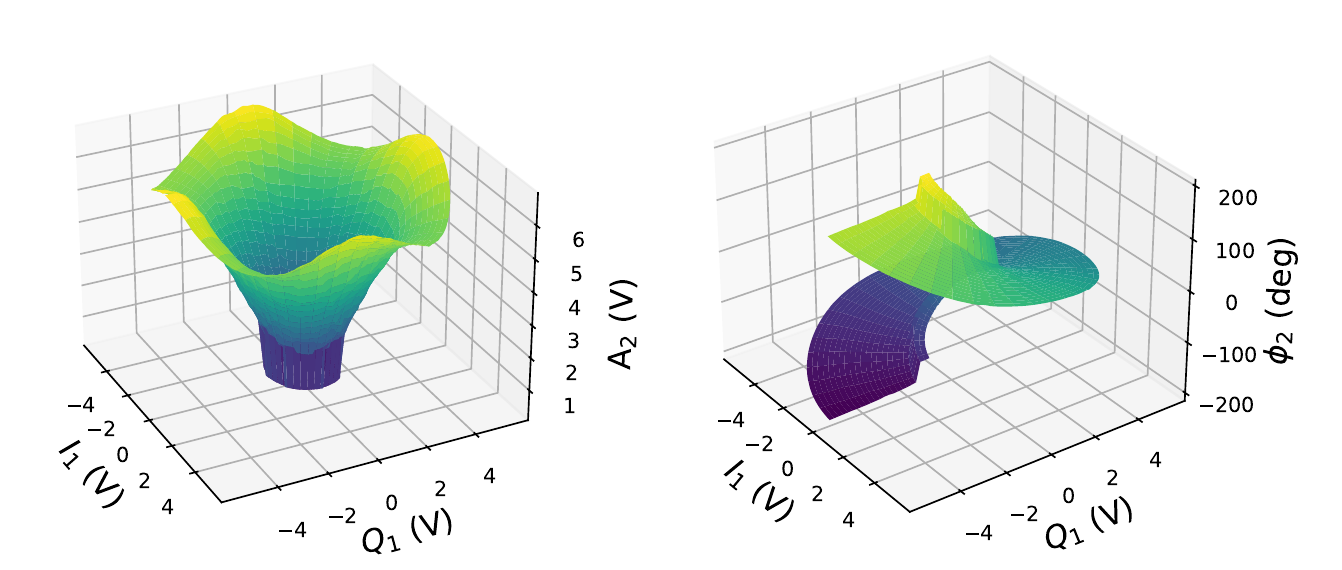}
    \caption{Map between the voltages applied to the modulation IQ mixer and the voltages measured on the demodulation IQ mixer. The demodulation IQ mixer voltages are expressed in terms of an amplitude and phase. In the ideal case, $A_2$ and $\phi_2$ would relate to $I_1$ and $Q_1$ through  Eq. \ref{eq:amp} and Eq. \ref{eq:phi} respectively. This empirical mapping corrects for deviations from the ideal.}
    \label{fig:mapping}
\end{figure*}

Let ($I_1$, $Q_1$) and ($I_2$, $Q_2$) denote the voltages on the modulation and demodulation mixers respectively. We can convert these voltages to amplitudes and phases through equations \ref{eq:amp} and \ref{eq:phi} to obtain modulation parameters ($A_1$, $\phi_1$), and the actual demodulated amplitude and phase in the cavity ($A_2$, $\phi_2$). For feedback on the RF phase and amplitude, we require a mapping from ($I_1$, $Q_1$) to ($A_2$, $\phi_2$) which will have a similar form to the analytic mapping with some non negligible deviations due to the non-linear response and thresholds of the Empower amplifier. To correct for these, we use a lookup table approach. The demodulated values of amplitude and phase are measured for a dense grid of input modulation parameters; this discrete mapping is then interpolated to create a continuous function $L:$ ($I_1$, $Q_1$)  $\rightarrow$ ($A_2$, $\phi_2$). An example of this mapping is shown in Fig. \ref{fig:mapping} and clearly shows an approximate correspondence to the expected conical and arc-tangent behavior.

A LabJack U3-HV with a LJTick-DAC is used to digitally read and write the IQ mixers' DC voltages; this device provides a -10 to 10 V range. The read and write DC voltages for the IQ mixers are typically low (around 0.5 V). Voltage dividers and low noise amplifiers [Analog Devices,  OP27] are used to boost the signal to fill the dynamic range of the LabJack. For each measurement in the feedback system, 100 individual digital measurements are performed and averaged to increase accuracy.

\begin{figure*}[ht]
    \centering
    \includegraphics[scale=0.67]{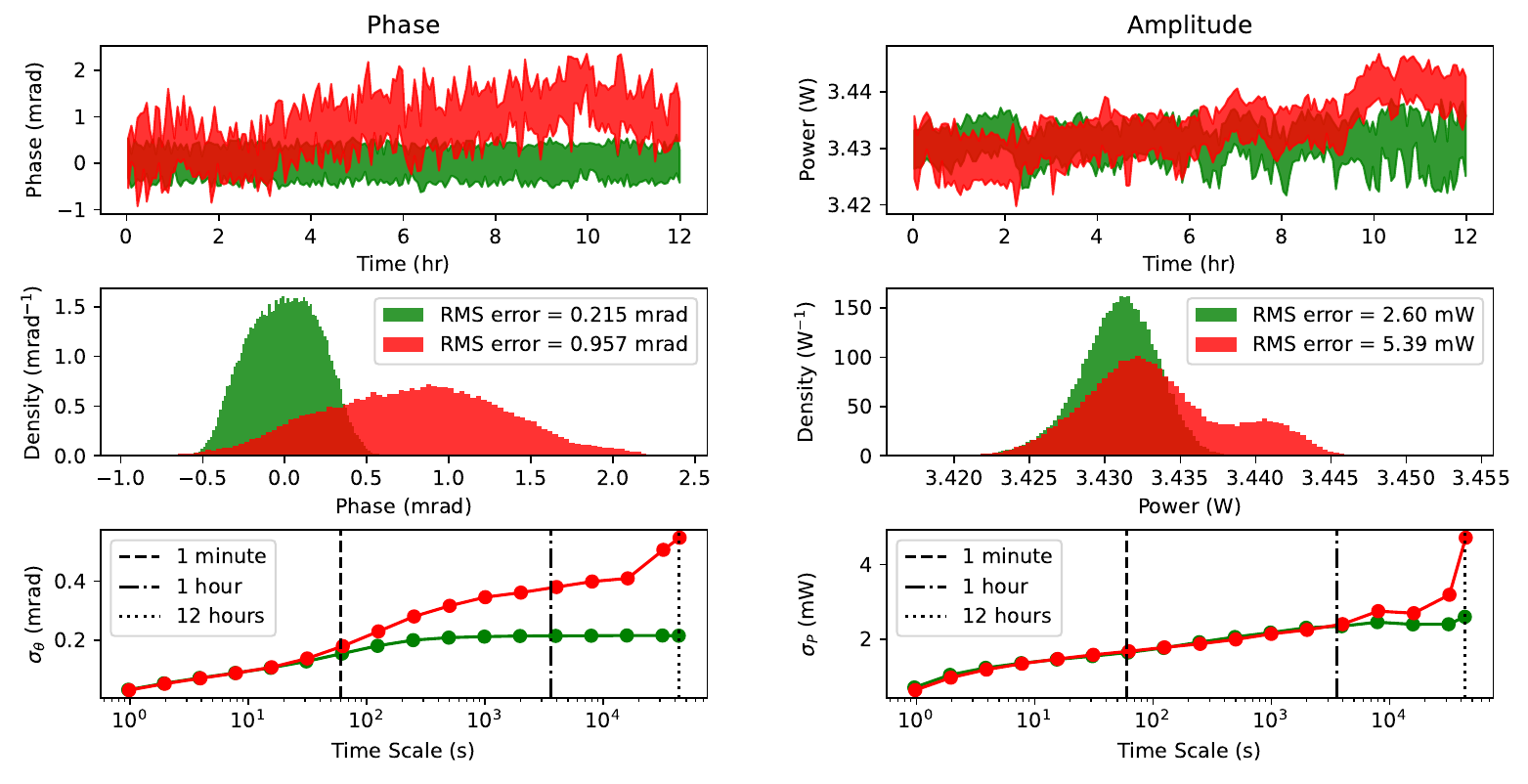}
    \caption{RF phase (left) and amplitude (right) stability with PID feedback on (green) and off (red). The top panels show the raw readings over a 12 hour period. The middle panels give histograms of these readings demonstrating the superior stability attained with feedback. The bottom panels give the averages of standard deviations of the raw data binned over different time scales.}
    \label{fig:feedback_perf}
\end{figure*}

UED experiments at the GARUDA beamline frequently require 12 hours of data acquisition to achieve a high quality signal-to-noise ratio. During this time, the same UED trace is scanned over many times and the results are averaged. Fig.~\ref{fig:feedback_perf} displays RF amplitude and phase data over two separate 12 hour periods to demonstrate the difference in RF stability with and without active feedback. It is apparent that activation of the vectorial feedback results in a significant improvement of phase stability on timescales as short as one hour while the improvement in amplitude stability (more than a factor of two) takes longer to appreciate. We can also compare the amplitude stability of the full vectorial feedback to a system where only phase feedback is applied. In this scenario, the phase feedback introduces significant additional amplitude jitter due to phase-amplitude correlations in RF phase shifter circuit elements \cite{siwick:jitter}. For the IQ mixer based design presented here, a phase only feedback system adds noise into the amplitude channel due to the cross-talk in the lookup table $L$. This can increase the amplitude drift by a factor of 100. Thus, simultaneous feedback for both quantities is essential for best performance when using these mixers.

Given the measured performances of the feedback loop shown in Fig. \ref{fig:feedback_perf}, we can estimate for this data set that the time-zero jitter due to fluctuations in RF cavity phase and power is lower than 20~fs RMS at the zero-field crossing. The total time-zero jitter in a UED experiment is a combination of this RF cavity jitter along with the timing jitter in the DRO and laser oscillator. This quantity is $\sigma_{\text{jitter}} \approx 50$ fs RMS.

\section{System Characterization}

We characterize the GARUDA beamline performance through diffraction measurements on a mono-crystalline sample of $1T$-TaS$_2$. This material is a widely studied quasi-two-dimensional compound belonging to the transition metal dichalcogenide family of materials. It is metallic at high temperatures ($>$550K), but below this temperature, it hosts a series of charge-density-wave (CDW) states possessing different characteristic wavevectors~\cite{di-salvo:bell,fazekas:crip}. The sample studied here was sectioned to approximately 60~nm thickness via ultramicrotomy and mounted freestanding on a copper TEM grid.

\begin{figure}
    \centering
    \includegraphics[scale=0.5]{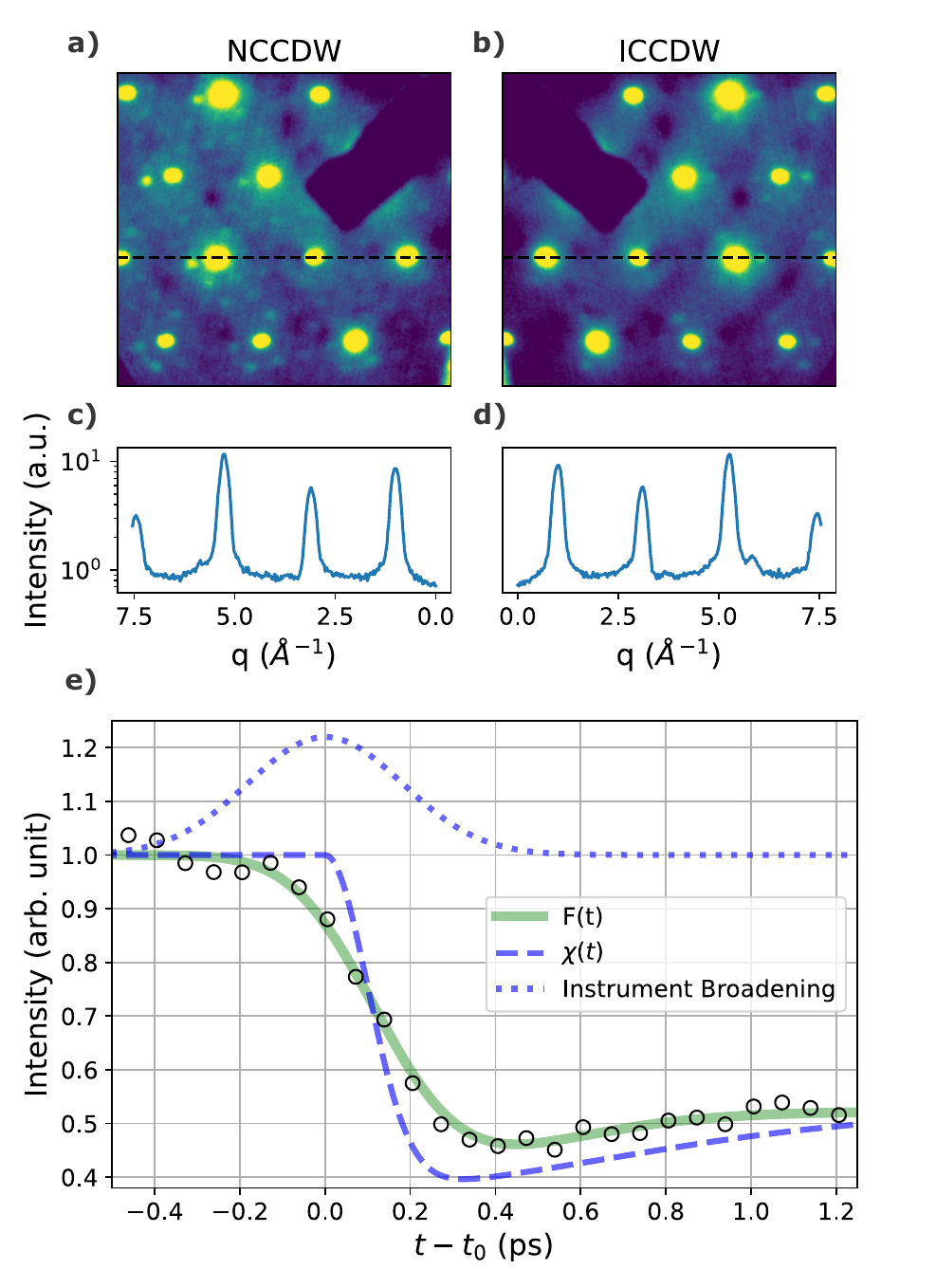}
    \caption{Electron diffraction patterns from $1T$-TaS$_2$ in the (a) nearly commensurate charge density wave (NCCDW) state and (b) incommensurate charge density wave (ICCDW) state. (c) and (d) Intensity cuts along the dashed lines drawn in (a) and (b). The average Bragg peak width (RMS) is measured to be $\sigma_q = $ 0.104 \AA$^{-1}$. (e) A UED time trace of the NCCDW peak intensity during a photo-induced transition to the ICCDW state. The data is fit to $F(t)$ (defined in Eq.~\ref{eq:F}) which is plotted as the solid green curve. The dashed blue curve is the intrinsic sample response given by $\chi(t)$ (defined in Eq.~\ref{eq:chi}). The dotted blue curve is the instrument response function which fits to an RMS width of $144 \pm 19$~fs. This data was collected with a pump fluence of 2.7~mJ/cm$^2$, 180~fs FWHM laser pulse width, 500~Hz repetition rate, and a bunch charge of 3.3~fC. RF power jitter was below 3 mW RMS. }
    \label{fig:timetrace}
\end{figure}

The momentum resolution of a diffraction instrument can be directly determined by analyzing Bragg peak widths collected from a known sample (1$T$-TaS$_2$). From the data in Fig. \ref{fig:timetrace}, we measure an RMS momentum resolution of $\sigma_q = $ 0.104 \AA$^{-1}$. This corresponds to a transverse coherence length of 9.62 \AA. This quantity is related to the RMS beam angular divergence at the sample ($\sigma_\theta$) according to,
\begin{align*}
\sigma_q = \frac{m\gamma\beta c}{\hbar} \sigma_\theta
\end{align*}
Thus, we have a divergence at the sample plane of 0.988~mrad. Additionally, the measured RMS beam transverse size at the sample is 37~$\mu$m (with the 100~$\mu$m pinhole inserted). Hence, the normalized beam emittance is approximately 11.6~nm$\cdot$rad. However, this measurement is not obtained exactly at the transverse focal point of the beam; therefore, the quoted value serves as an upper bound on the normalized emittance. 

The temporal resolution of the system is characterized through measurements of CDW peak dynamics in 1$T$-TaS$_2$. Specifically, we present UED measurements on the photo-induced transition from the room temperature nearly-commensurate charge-density-wave (NCCDW) to the incommensurate charge-density wave (ICCDW) phase. All measurements are performed at 500 Hz repetition rate with a laser pump of 835 nm wavelenth and 750 $\mu$m RMS transverse width.  

Time-resolved X-ray diffraction studies on this transition have shown a complete suppression of the NCCDW peaks within $\sim$250 fs of pump excitation followed by a slight recovery~\cite{Laulhe:trxrd}. These data provide a measure of the intrinsic response time of the photo-induced NCCDW-to-ICCDW transition. We utilize a phenomenological function as an effective Green’s function, $\chi(t)$, that describes the intensity of the NCCDW diffraction peak as a function of time in the limit of zero instrument broadening.
\begin{align}
\begin{split}
    \chi(t) &=
        \begin{cases}
        1 & \text{if } t < 0 \\
        1 + I_D f_1(t) + (I_{\infty} + I_D - 1) f_2(t) & \text{if } t > 0
        \end{cases} \\
    f_1(t) & = -1 + e^{-(t/\tau_1)^2} \\
    f_2(t) & = 1 - e^{-(t/\tau_2)^2} \\
\end{split}
\label{eq:chi}
\end{align}
Here $I_D$ defines the magnitude of the initial intensity drop, and $I_{\infty}$ is the intensity in the limit of $t \rightarrow \infty$. The intrinsic response time was previously measured using time-resolved X-ray diffraction by Laulh\'e et al. whose results fit to $\tau_1 = 140$ fs, a timescale that characterizes the initial peak intensity drop. In order to quantify our instrument response function, we hold $\tau_1$ fixed at this value. The peak intensity recovers on a timescale that is characterized by the variable $\tau_2$. 

Our data is fit to a convolution of this Green’s function with a Gaussian of variance $\tau_i^2$ to represent the instrument broadening. Thus, the fit curve is given by, 
\begin{align}
F(t) = \chi(t) * \frac{1}{\sqrt{2\pi\tau_i^2}} \exp{\left(-\frac{t^2}{2\tau_i^2}\right)}
\label{eq:F}
\end{align}
Fig.~\ref{fig:timetrace} shows a fit of $F(t)$ to the UED data on the photo-induced NCCDW to ICCDW transition. The data was collected at optimized RF power and RF phase; thus, the value of $\tau_i$ extracted from this curve is the minimum achieved RMS instrument response time: $144 \pm 19$ fs. The quantity $\tau_i$ contains all sources of instrument broadening; however, the width of the IR pump ($\sigma_{\text{pump}} = 85$ fs RMS), the width of the electron probe pulse ($\sigma_t$), and the arrival time jitter ($\sigma_{\text{jitter}} \approx 50$ fs RMS) are the dominant factors. By subtracting in quadrature the length of the laser pump pulse and the contribution of the timing jitter, we obtain an estimate for the electron bunch RMS temporal width $\sigma_t = 105$~fs.

Limitations on this temporal width at focus are primarily due to space-charge effects and phase-space non-linearity \cite{cropp2023virtual}. Phase-space non-linearity refers to a deviation from perfect correlation between particle position and velocity along the beam axis; this will cause the negative, linear chirp applied by the RF cavity to imperfectly focus the bunch. Improvements here can be obtained by either increasing the electron beam energy or decreasing the distance between the RF lens and the sample plane. 


\begin{figure}
    \centering
  \includegraphics[scale=0.48]{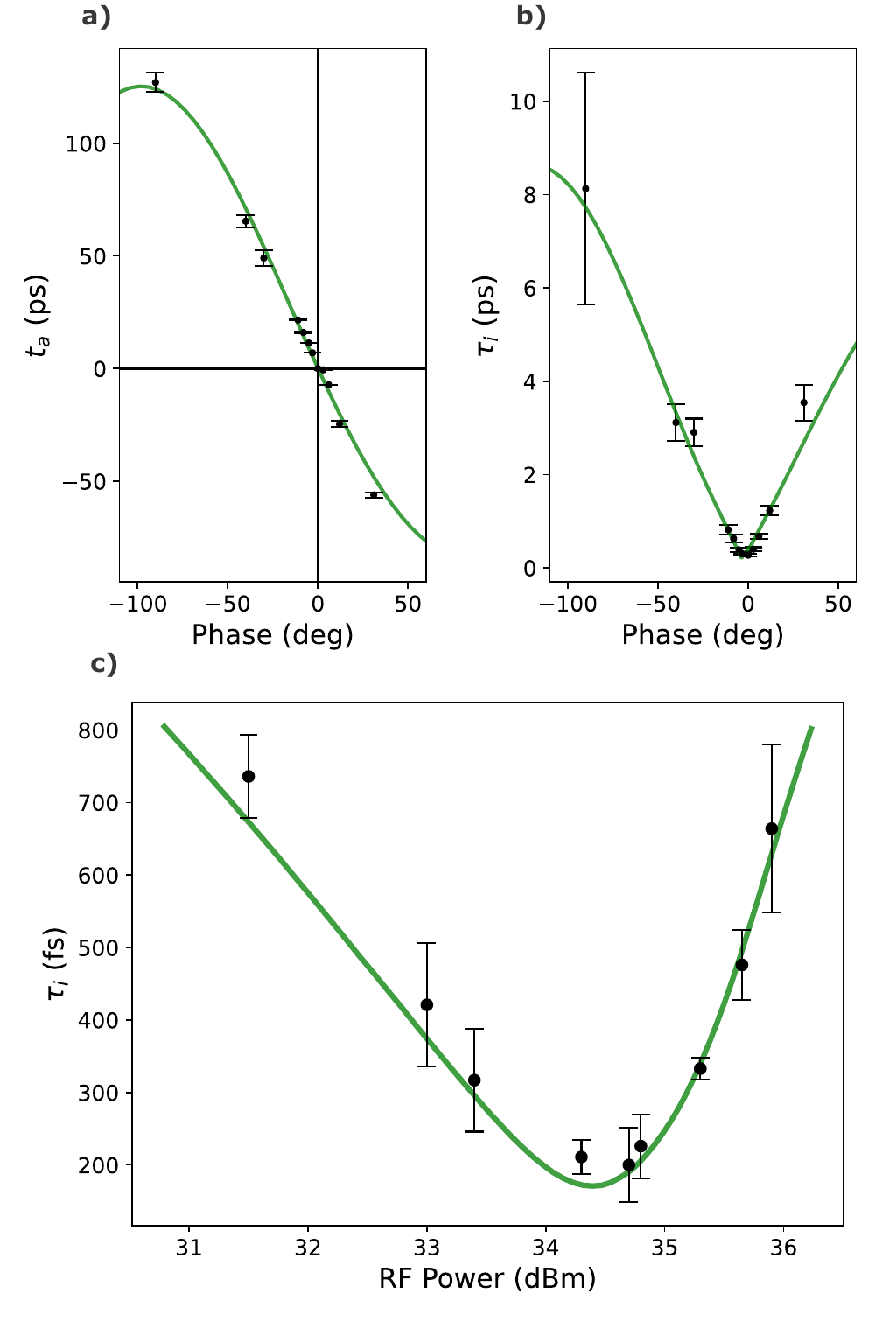}
    \caption{(a) Electron bunch relative arrival time ($t_a$) as a function of RF phase. The slope of $t_a$ vs. phase at the zero-crossing is 1.54 ps/$^\circ$. (b) Instrument response time ($\tau_i$) as a function of RF phase. The data in both (a) and (b) were collected for an RF power of 34.5 dBm. (c) Instrument response time as a function of RF cavity power at zero phase. The green solid lines are from the GPT model. Each data point in these plots is extracted from a UED time-trace with pump fluence of 4 mJ/cm$^2$, 180~fs FWHM laser pulse width, 500~Hz repetition rate, and bunch charge of 1 fC. Note that this data was collected before optimized RF amplitude stability was achieved; here the power jitter was at the level of 200 mW RMS.}
    \label{fig:ampphasecomp}
\end{figure}

To empirically determine the zero-crossing compression phase of the RF cavity, one can measure a set of Bragg peak positions as a function of RF phase at fixed RF power - shown in Fig. \ref{fig:deltaQ}. This curve shows the magnification factor of the initial diffraction pattern due to RF acceleration from the cavity. If there is no magnification (factor is equal to 1), then zero net momentum was imparted to the electron beam; this can occur for either a compression phase or an expansion phase that are separated by 90$^\circ$. To differentiate the two, it is necessary to know whether increasing phase leads or lags the RF wave. In our case, phase leads the wave which means that the zero-crossing where ${dt_a}/{d\psi}|_{\psi=0} < 0$ is the compression phase. The precision of this method is limited by the position resolution of the Bragg peaks in the diffraction pattern. Using Gaussian fits for the peak profiles allows sub-pixel accuracy in the determination of peak centroids; nevertheless, the measurement of the compression phase is still uncertain to within a few degrees with this method.

Significantly better precision on the compression phase determination is obtained from time-zero measurements in a UED experiment. A reference case time-zero is measured from a time-trace in which no power is supplied to the RF cavity. This defines the zero of the electron arrival time ($t_a = 0$). Once power is supplied to the RF cavity, deviation from the zero-crossing phase will result in a non-zero $t_a$. A UED time-trace measures this quantity to within 100 fs error (1$\sigma$ confidence) which allows the RF phase to be tuned to the zero-crossing with 1.1 mrad precision.

Fig.~\ref{fig:ampphasecomp}(a) shows results from UED data collected at a fixed RF power for a number of different RF phases. The phase where $t_a=0$ is a point at which zero net momentum is imparted to the electron beam. 
The slope of $t_a(\psi)$ at the zero-crossing is measured to be -1.54 ps/$^\circ$ in good agreement with Eq.~\ref{dtdpsi}.

The dependence of the system response time ($\tau_i$) on phase at the ideal RF power is shown in Fig.~\ref{fig:ampphasecomp}(b). The range of compressing phases at the ideal power is quite large; however, the onset of time-zero jitter due to RF power fluctuations can quickly increase the system response time as phase is moved away from the zero-field crossing. For the data presented in Fig.~\ref{fig:ampphasecomp}, the total RF power jitter was $\sim$200~mW RMS; however, the phase jitter was held below 0.5~mrad. Even with this comparatively high jitter in RF power, the system response time at the zero-field crossing is still below 200~fs. As shown in Figure \ref{fig:feedback_perf}, the RF power jitter has since been improved to the 2.6~mW level. 

Fig.~\ref{fig:ampphasecomp}(c) displays measurements of the system response time at various RF powers while the phase is set at the zero-crossing. This corresponds to varying the focal length of the RF longitudinal lens. Clearly, a power that is too high (low) yields a focus before (after) the sample plane. This data is found to be in agreement with the expected curve from particle tracking simulations and shows a relatively large, approximately 300 mW range, for an acceptable temporal focus below 200~fs RMS.

\section{Conclusions}
In conclusion, a novel UED beamline operating at 40.4~keV electron energy has been commissioned at UCLA based on RF compression to obtain an RMS temporal resolution of 144$\pm$19~fs. The cavity is operated in CW mode which allows for fast feedback control. A vector-based feedback loop using IQ mixers to read and control the amplitude and phase of the cavity RF field is implemented. This active stabilization of both amplitude and phase allows for new operating conditions for RF compression. For instance, compression away from the zero-field crossing for tuning of the electron energy or RF focusing can be employed using the vector-based feedback scheme. Additionally, implementation of two amplitude stable RF cavities could be used for precise energy collimation of the electron bunch, which would enable high precision momentum resolved electron energy loss spectroscopy (EELS)~\cite{duncan2020}. Further improvements in temporal resolution can be obtained by increasing the energy from the electron gun (currently limited by arcing at the cathode) and optimizing the compression for different bunch charges. Compression of the pump laser pulse and reduction of the RF cavity-to-sample distance would also improve the instrument response time. Further advances are possible in which timing jitter is handled through post processing. Since the UED scans are performed stroboscopically, good resolution electronic measurements could enable a scheme in which each probe shot is assigned its own arrival time and machine learning/artificial intelligence is used to compensate for the residual timing jitter~\cite{cropp2023virtual}. 

\appendix

\nocite{*}

\acknowledgements
This work was supported by the U.S. Department of Energy (DOE), Office of Science, Office of Basic Energy Sciences under Award No. DE-SC0023017, the National Science Foundation under Grant No. DMR-1548924 and NSF MRI grant 1828705. We would also like to thank Nathan Burger, Amir Amhaz, Kirill Talesky, Alexander Ody, Shiying Wang, Mikhael Rasiah and Jayanti Higgins for their help on various aspects of the beamline construction.

The data that support the findings of this study are available from the corresponding author upon reasonable request.

\bibliography{rf}

\end{document}